\documentclass[twocolumn,showpacs,preprintnumbers,amsmath,amssymb]{revtex4}
\usepackage{graphicx}
\usepackage{dcolumn}
\usepackage{bm}
\usepackage{epsf}

\newcommand{\bea}{\begin{eqnarray}}
\newcommand{\eea}{\end{eqnarray}}
\newcommand{\nea}{\nonumber\end{eqnarray}}
\newcommand{\eq}[1]{eq.~(\ref{#1})}

\newcommand{\Eq}[1]{Eq.~(\ref{#1})}

\newcommand{\ur}[1]{(\ref{#1})}

\newcommand{\beq}{\begin{equation}}
\newcommand{\eeq}{\end{equation}}

\newcommand{\la}[1]{\label{#1}}
\newcommand{\ba}{\begin{array}}
\newcommand{\ea}{\end{array}}

\newcommand{\at}{\overline{10}}

\newcommand{\noi}{\noindent}
\newcommand{\nn}{\nonumber}

 \def\Dirac#1{#1\hskip-6pt/}
 \def\dd{\Dirac\partial}

\begin{document}

\title{From Pions to Pentaquarks~\cite{F1}}
\author{Dmitri Diakonov$^{a,b,c}$}

\vskip 0.3true cm

\affiliation{
$^a$ Thomas Jefferson National Accelerator Facility, Newport News, VA 23606, USA\\
$^b$ NORDITA, Blegdamsvej 17, DK-2100 Copenhagen, Denmark\\
$^c$ St. Petersburg Nuclear Physics Institute, Gatchina, 188 300, St. Petersburg, Russia
}

\date{June 3, 2004}

\begin{abstract}
I overview the physical picture of QCD at low energies that has led to the prediction 
of a narrow exotic baryon $\Theta^+$ which cannot be made of three quarks. 
The very narrow width of the $\Theta^+$ and a possible reason why it is seen in low- but not 
in high-energy experiments are briefly discussed. 
\end{abstract}

           

\maketitle

\section{Problems in the constituent quark models}

Everyone knows that nucleons are ``made of three quarks'' but there have long been 
problems with this simplistic view. Let me list just three difficulties out of 
many.

Already in the 1970's it became known that there were many antiquarks in nucleons,
seen explicitly in the deep inelastic $\bar\nu$ scattering and in the Drell--Yan 
lepton pair production. The amount and the $x$-distribution of antiquarks could not 
be explained by simple perturbative bremsstrahlung: some non-perturbative amount 
of antiquarks must have been there at low momenta~\cite{DDT}. All modern fits to deep 
inelastic scattering data include a sizeable antiquark distribution at low virtuality, 
e.g. Ref.~\cite{GRV}. 

By measuring the polarized quark distributions in the nucleon it was shown in the 
1980's that only a small fraction of the nucleon spin was carried by the quarks' spin 
(including possible antiquarks). The modern value of this fraction is $0.3\pm 0.1$~\cite{FJ}. 
The rest of the nucleon angular momentum can be carried by gluons or by some orbital momentum. 
The paradox, named the ``spin crisis'', sharpens when one realizes that gluons 
couple to the measured polarized distribution function in the higher order in the
$\alpha_s$ coupling and therefore cannot contribute much, whereas the three
constituent quarks are believed to be in the $s$-state and hence carrying no 
orbital momentum.  

Known least, but probably the worst departure of the constituent quark model
from reality is in the value of the so-called nucleon sigma term. It is experimentally
measured in low-energy $\pi N$ scattering, and its definition is the scalar quark 
density in the nucleon, multiplied by the current (or bare) quark masses,
\beq
\sigma=\frac{m_u+m_d}{2}\,<N|\bar u u+\bar d d|N>=67\pm 6\,{\rm MeV}.
\la{sigma_term}\eeq
I have given here the modern value~\cite{Arndt1}; the historical change
in this very important quantity as well as the uncertainties in extracting
it from the raw data have been recently discussed in Ref.~\cite{Schweitzer}.
The standard values of the current quark masses are~\cite{Leutwyler}  
$m_u\simeq 4\,{\rm MeV},\,m_d\simeq 7\,{\rm MeV}$ (and $m_s\simeq 150\,{\rm MeV}$).
In the non-relativistic limit, the scalar density is the same as the
vector density; therefore, in this limit the matrix element above is
just the number of $u,d$ quarks in the nucleon, equal to 3. If $u,d$
quarks are relativistic, the matrix element is less than three. Hence,
in the constituent quark model
\beq
\sigma_{\rm quarks}\leq \frac{4\,{\rm MeV}+7\,{\rm MeV}}{2}*3
=17.5\,{\rm MeV},
\la{sigma_quarks}\eeq 
i.e. {\it four} times less than experimentally!    
 
The $\sigma$ term has the physical meaning of the part of the nucleon mass
which is due to current quark masses. On the one hand $\sigma$-term is too big
from the 3-quark point of view but on the other hand, it is too small as 
it explains only 7\% of the nucleon mass, or even only 2\% if we use
the quark estimate \ur{sigma_quarks}. Where does the rest 93\% of the nucleon 
mass come from?

I would like to stress that the situation here is radically different
from all other bound states we know. Be it atoms or nuclei, their masses are
always {\it less} than the appropriate ``$\sigma$-term'' or the sum of their
constituents' masses. This paradox has to be explained first of all.

\section{Origin of hadron masses}

The most important happening in QCD from the point of view of the light hadron
structure is the Spontaneous Chiral Symmetry Breaking: as its result, almost 
massless $u,d,s$ quarks get the dynamical momentum-dependent masses $M_{u,d,s}(p)$, 
and the pseudoscalar mesons $\pi, K, \eta$ become light (pseudo) Goldstone bosons.
 
This story starts from the gluon vacuum. In the absence of any matter and 
sources, i.e. in the vacuum, the gluon field experiences zero-point quantum 
fluctuations whose snapshot can be visualized in lattice simulations (Fig.~1, top).
These are the normal quantum fluctuations, experienced also by the electromagnetic 
field. What is peculiar for the non-Abelian gluon field, is that beneath
zero-point oscillations there are specific large fluctuations called
instantons (for a recent review see Ref.~\cite{inst_at_work}). One can measure
their average size and the average separation between the peaks, which appear 
to be roughly $1/3\,{\rm fm}$ and $1\,{\rm fm}$, respectively. These key
numbers had been suggested by Shuryak~\cite{Shuryak} and obtained from 
$\Lambda_{\overline{\rm MS}}$ \cite{DP1} long before lattice measurements
became available.  

\begin{figure}[]
\epsfxsize=19pc 
\epsfbox{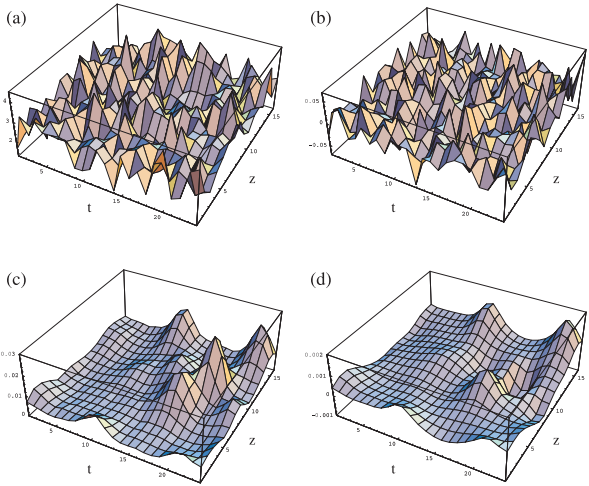} 
\caption{A snapshot of quantum fluctuations of the gluon field in
the vacuum~\cite{CGHN}. ``Cooling" the normal zero-point oscillations (top) 
reveals that they lie on top of large gluon fluctuations, which are 
identified with instantons and anti-instantons with random positions 
and sizes (bottom). The left column shows the action density and the 
right column shows the topological charge density for the same snapshot,
where instantons are peaks and anti-instantons are holes.}
\vskip -0.5true cm
\end{figure}

\begin{figure}[b]
\epsfxsize=19pc 
\epsfbox{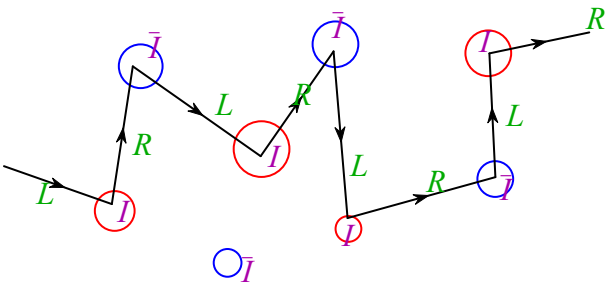} 
\caption{Propagation of the originally massless quarks through a random instanton
ensemble renders them a dynamical mass $M(p)$~\cite{DP2}.}
\end{figure}

Now we switch in light quarks and ask how quarks propagate through this 
fluctuating gluon vacuum (Fig.~2). The result~\cite{DP2} is that the nearly massless 
quarks gain a large dynamical mass $M(p)$ plotted in Fig.~3. This is in fact
a beautiful and a non-trivial phenomenon. The point is, massless quarks generally conserve
their helicity when interacting with gluons. Therefore, the mass is not generated
in any order in perturbation theory. In other words, normal zero-point
oscillations of the gluon field depicted on the top of Fig.~1 do not generate the
quark mass. Only the large instanton fluctuations shown in the bottom of 
Fig.~1 have the power of flipping quark helicity. It is a non-perturbative effect.
Quarks ``hop" from one randomly situated instanton fluctuation to another, 
each time flipping the helicity. As a result they get a dynamical mass. 
It goes to zero at large momenta since quarks with very high momenta are not 
affected by the background, even if it is strong gluon field as in the case of instantons.

\begin{figure}[t]
\epsfxsize=16pc 
\epsfbox{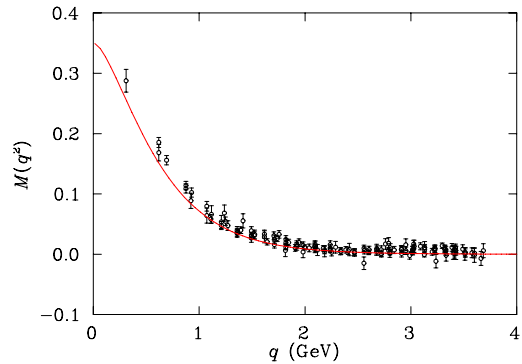} 
\caption{Dynamical quark mass $M(p)$ from a lattice simulation~\cite{Mlat}. Solid curve: 
from instantons, {\bf no fitting}~\cite{DP2}.}
\end{figure}

Another striking confirmation of the instanton mechanism of the spontaneous chiral 
symmetry breaking~\cite{DP2} came recently from another lattice study~\cite{Faccioli}. 
In Fig.~4 the ratio of the helicity-flip to non-flip correlation functions of flavor 
non-singlet currents is plotted. Random instantons explain very precisely the
non-trivial shape of this ratio as function of the distance.

\begin{figure}[t]
\epsfxsize=13pc 
\epsfbox{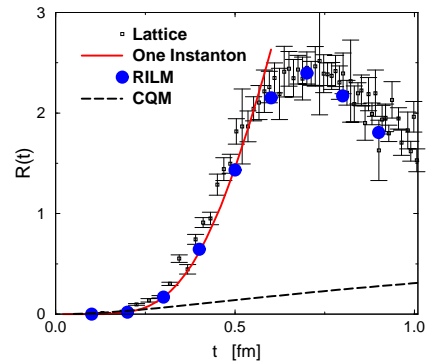} 
\caption{The ratio of the helicity-flip to the non-flip correlation functions 
as function of distance, measured in lattice simulations, and compared to the 
instanton calculation (circles), {\bf no fitting}~\cite{Faccioli}.} 
\vskip -0.5true cm
\end{figure}

The dynamical mass plotted in Fig.~3 is key to understanding light hadrons' properties.
Its value at zero momentum,
\beq
M(0)\simeq 350\,{\rm MeV},
\la{Mof0}\eeq
is what is usually called the ``constituent" quark mass (to be distinguished from the
small input ``current" quark masses of the QCD lagrangian.) The half-width of $M(p)$
corresponds roughly to the size of the constituent quarks, 
\beq
\rho\simeq \frac{1}{600\,{\rm MeV}}=\frac{1}{3}\,{\rm fm},
\la{size1}\eeq
In the instanton approach, these numbers are calculated from the average separation 
between and the average size of the instanton fluctuations which, in their turn, can be 
computed from the value of $\Lambda_{\overline{\rm MS}}$~\cite{DP1}.

By the way, there is no real solution of the equation $p^2=M^2(-p^2)$, meaning that quarks
cannot be observable, only their bound states! 

From the above numbers one can immediately estimate hadron sizes and masses. 

\underline{Hadron sizes} are determined, via the uncertainty principle, 
by the inverse $M(0)$, or are somewhat larger if constituent quark binding is loose:  
\beq
R\geq \frac{1}{M} \sim 0.8\;{\rm fm}. 
\la{size2}\eeq
It is crucial that hadron sizes are significantly larger than quark sizes, otherwise
the constituent quark ideas would never have worked.

\underline{Hadron masses} are roughly twice the constituent quark mass for
the typical vector mesons, and thrice $M$ for baryons: 
\bea
\la{vector_mass}
{\rm Vector\; mesons:}\; m_V&\approx &2M\simeq 700\,{\rm MeV},\\
\la{baryon_mass1}
{\rm 3\!-\!quark\; baryons:}\; m_B&\approx &3M\simeq 1000\,{\rm MeV}.
\eea
In pseudoscalar mesons, however, two constituent quark masses are completely
`eaten up' by strong interaction, which is guaranteed by the Goldstone theorem
stating that once chiral symmetry of strong interactions is spontaneously
broken, pseudoscalar mesons must be very light. Their masses are fully determined 
not by the constituent but by the current quark masses, in accordance 
with the Gell-Mann--Oakes--Renner formula~\cite{Leutwyler}: 
\beq
m_P^2=(M+M-2M)^2+\frac{m_q<\!\bar q q\!>}{F_\pi^2}.
\la{pseudo_mass}\eeq
The very strong interactions between quarks in the pseudoscalar channel is due
to the same forces that are responsible for the originally nearly massless quarks
obtaining a dynamical mass $M(p)$: they are due to instantons, see Fig.~5. 

\begin{figure}[]
\epsfxsize=12pc 
\epsfbox{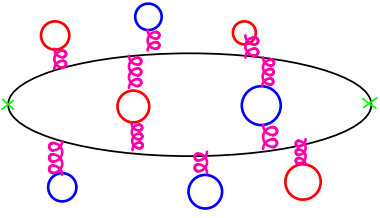} 
\caption{The same instantons that bring in $M(p)$ are responsible for very strong
binding of quarks into a nearly massless pion, which is guaranteed by the Goldstone 
theorem. Summation of diagrams demonstrating it: see Ref.~\cite{DP2}.}
\end{figure}

\section{What forces bind constituent quarks in baryons?}

The standard answer is: the linear confining potential, chromoelectric flux tubes. 
However, in the real world with very light pions it cannot be correct 
because it is energetically favorable to break an extended string, and produce 
pions, see Figs.~6,7.  

\begin{figure}[]
\epsfxsize=11pc 
\epsfbox{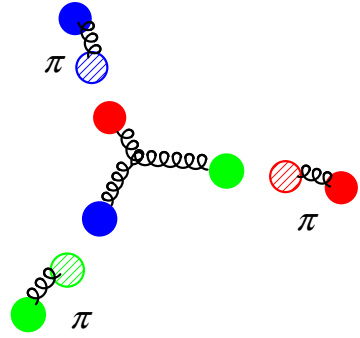} \caption{Confining strings whose energy is linearly
rising with quark separation break when their energy exceeds the threshold 
for meson production.}
\end{figure}

\begin{figure}[]
\epsfxsize=18pc 
\epsfbox{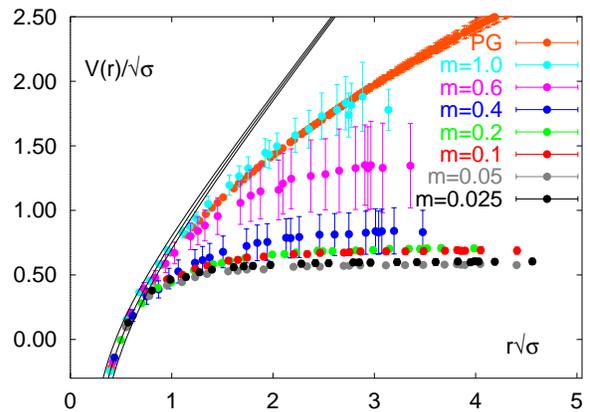} 
\caption{Reduction of the linear confining potential in the presence
of light quarks, from lattice simulations at high temperatures but below
the phase transition~\cite{Karsch}. The scales are $425\,{\rm MeV}/0.47\,{\rm fm}$.
See also Ref.~\cite{DeGrand} where lattice simulations were 
performed at zero temperature but larger pion mass.}
\vskip -0.3true cm
\end{figure}

In light baryons, it is sufficient to exceed the mass of one additional pion
to make the flux tube energetically unfavorable. Pion production makes
the color string break. Since pions are very light one can estimate that it 
happens at a very early $\simeq\! 0.3\,{\rm fm}$ separation between quarks, 
which is the size of the constituent quark itself! Meanwhile, the average separation
of quarks inside a baryon is much bigger -- typically about $0.8\,{\rm fm}$.
Therefore, color strings or color-electric flux tubes can hardly be responsible
for quark binding.  All what is left, are constituent quarks interacting with pions, 
but {\bf they themselves are made of constituent quarks}.

How to present this queer situation mathematically? There is actually not much freedom
here: the interaction of pseudoscalar mesons with constituent quarks is dictated
by chiral symmetry. It can be written in the following compact form~\cite{DP2}:

\beq 
{\cal L}_{\rm eff}=\bar q\;\left[i\dd-M\exp(i\,\gamma_5\,
\pi^A \lambda^A/F_\pi) \right]\,q,\quad \pi^A=\pi,K,\eta. 
\la{lagrangian}\eeq 
Because of the necessary exponent of the pseudoscalar field, it is an essentially
non-linear interaction. The exponent here is a $3\times 3$ flavor matrix acting on 
the 3-vector $(u,d,s)$. It is also a $4\times 4$ matrix acting on the Dirac 4-spinor
indices of quarks. 

[A note for experts: Since \eq{lagrangian} is an effective low-energy theory, 
one expects formfactors in the constituent quark -- pion interaction; in particular, 
as we know already $M(p)$ is momentum-dependent. \Eq{lagrangian} is written in the 
limit of zero momenta. A possible wave-function renormalization factor $Z(p)$ can be
also admitted but it can be absorbed into the definition of the quark field.
The low-energy theory similar to \eq{lagrangian} has been suggested in 1984 by several
groups~\cite{Georgi-Manohar,Birse-Banarjee,Kahana-Ripka}. However, in these 
influential papers an additional kinetic energy term for pions has been added to 
\eq{lagrangian}. It is neither necessary nor even seem to be correct: the kinetic 
energy term (and higher derivatives) for pions appears from integrating out quarks, or, 
in other words, from quark loops, see Fig.~8. It is in accordance with the fact that
pions are not `elementary' but a composite field, made of constituent quarks. In this
form, \eq{lagrangian} has been derived from instantons in Ref.~\cite{DP2}.]

\begin{figure}[]
\epsfxsize=13pc 
\epsfbox{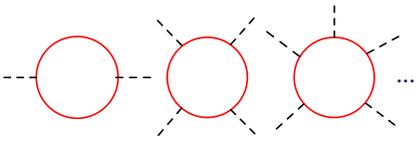} 
\caption{Pseudoscalar mesons are themselves bound states of constituent quarks:
they propagate and interact via virtual quark-antiquark pairs. The first diagram
is pion's propagation, the second one describes (correctly) the $s$- and $d$-wave 
pion scattering at low energies, the third gives the 5-prong process 
$K^+K^-\to \pi^+\pi^0\pi^-$ which is theoretically known from the Wess--Zumino term. 
The sum of all diagrams with any number of external legs is called the effective chiral
lagrangian.}
\end{figure}

Constituent $u,d,s$ quarks necessarily have to interact with the $\pi,K,\eta$ 
fields according to \eq{lagrangian}, and the dimensionless coupling constant is 
actually very large:
\beq 
g_{\pi qq}(0)=\frac{M(0)}{F_\pi}\simeq 4. 
\la{gpiqq}\eeq 
Alternatively, one can estimate this coupling as the pion-nucleon constant 
$g_{\pi NN}\simeq 13.3$ divided by three quarks, which gives approximately the same.

\section{What is the nucleon?}

The chiral interactions of constituent quarks in baryons, following from  
\eq{lagrangian}, are schematically shown in Fig.~9. 
Antiquarks are necessarily present in the nucleon
as pions propagate through quark loops. The non-linear effects in
the pion field are essential since the coupling is strong. I
would like to stress that this picture is a model-independent
consequence of the spontaneous chiral symmetry breaking. 
One cannot say that quarks get a constituent mass but throw away
their strong interaction with the pion field. In principle,
one has to add perturbative gluon exchange on top of Fig.~9. However,
$\alpha_s$ is never really strong, such that gluon exchange can
be disregarded in the first approximation. The large value of
the pion-quark coupling \ur{gpiqq} suggests that Fig.~9
may well represent the most essential forces inside baryons. 

\begin{figure}[]
\epsfxsize=16pc 
\epsfbox{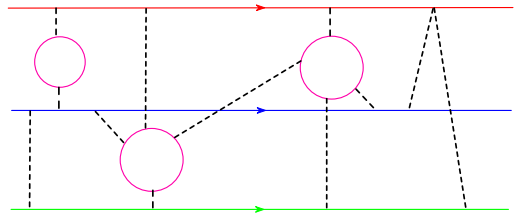} 
\caption{Quarks in the nucleon (solid lines), interacting via pion
fields (dash lines).}
\vskip -0.5true cm
\end{figure}

The low-momenta effective theory defined by \eq{lagrangian} is a big
step forward, as compared to the original formulation of QCD: it operates
with the adequate degrees of freedom, namely the dynamically-massive quarks and
the light pseudoscalar meson fields, relevant at low energies. The transition
to these new degrees of freedom is similar to the transition from
QED (= the microscopic theory of the atoms) to the electrons in a material,
whose mass is not the original $0.511\;{\rm MeV}$ but a heavier effective one,
and whose most important interaction at the atomic ``low energies''
is not the Coulomb (read: gluon) but rather the phonon (read: pion) exchange.
Phonons are collective excitations of atomic lattices, and they are
Goldstone bosons associated with the spontaneous breaking of the translational
symmetry by the lattice. Pions are collective excitations of the QCD
ground state, and they are Goldstone bosons associated with the spontaneous
breaking of chiral symmetry. Their interaction with quarks which
obtained the dynamical mass also owing to the same spontaneous chiral symmetry
breaking, is dictated by symmetry. Only the parameters of the
low-energy effective action \ur{lagrangian} have to be determined from a microscopic 
theory -- in this case instantons seem to do this job very well. 

Continuing the analogy, the Cooper pairing of electrons in a superconductor 
is due to phonon exchange which is much stronger than the Coulomb force 
between electrons (being actually a repulsion). As we shall
see in a moment, the binding of quarks into a nucleon can be explained
as due to their interaction with pions. Although the low-momenta
effective theory \ur{lagrangian} is a great simplification as compared
to the microscopic QCD, it is still a strong-coupling relativistic quantum field
theory. Summing up all interactions inside the nucleon of the kind shown
in Fig.~9 is a difficult task. Nevertheless, it can be solved 
exactly in the limit of large number of colors $N_c$. It is widely 
believed~\cite{tH2,W1} that the real world with $N_c=3$ is 
not qualitatively different from the imaginary world at $N_c\to\infty$. 
For $N_c$ colors the number of constituent quarks in a baryon is $N_c$, 
and all quark loop contributions in Fig.~9 are also proportional to $N_c$. 
Therefore at large $N_c$, quarks inside the nucleon produce a large, nearly 
classical pion field: quantum fluctuations about the mean field are 
suppressed as $1/N_c$. The same field binds the quarks; therefore it is 
called the {\em self-consistent} field. [A similar idea is exploited in the 
shell model for nuclei.] The problem of summing up all diagrams of the type 
shown in Fig.~9 is thus reduced to finding a classical self-consistent pion 
field. As long as $1/N_c$ corrections to the mean field results are under
control, one can use the large-$N_c$ logic and put $N_c$ to its
real-world value 3 at the end of the calculations. 

The model of baryons based on these approximations has been
named the Chiral Quark Soliton Model (CQSM)~\cite{DP-CQSM}. The
``soliton'' is another word for the self-consistent pion field in
the nucleon. However, the model operates with explicit
quark degrees of freedom, which enables one to compute any type of
observables, {\it e.g.} relativistic quark (and antiquark!)
distributions inside nucleons~\cite{SF}, and the quark light-cone
wave functions~\cite{light-cone}. In contrast to the naive quark models,
the CQSM is relativistic-invariant. Being such, it necessarily
incorporates quark-antiquark admixtures to the nucleon. Quark-antiquark
pairs appear in the nucleon on top of the three valence quarks
either as particle-hole excitations of the Dirac sea (read: mesons) or as
collective excitations of the mean chiral field.

\begin{figure}[]
\epsfxsize=15pc 
\epsfbox{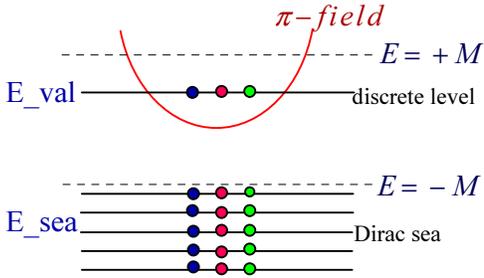} 
\caption{If the trial pion field is large enough (shown schematically by
the solid curve), there is a discrete bound-state level for three `valence'
quarks, $E_{\rm val}$. One has also to fill in the negative-energy
Dirac sea of quarks (in the absence of the trial pion field it corresponds
to the vacuum). The continuous spectrum of the negative-energy levels is
shifted in the trial pion field, its aggregate energy, as compared to the
free case, being $E_{\rm sea}$. The nucleon mass is the sum
of the `valence' and `sea' energies, multiplied by three
colors, ${\cal M}_N=3\left(E_{\rm val}[\pi(x)]+E_{\rm sea}[\pi(x)]\right)$.
The self-consistent pion field binding quarks is the one minimizing the nucleon
mass.}
\vskip -0.4true cm
\end{figure}

It should be stressed thus that there is nothing odd or dramatic
in calling the nucleon a chiral soliton. It is no more soliton
than an atom which, at large $Z$, can be well described by the
Thomas--Fermi method where one considers $Z$ electrons in the
self-consistent field, in that case it is the electrostatic field.
One can view an atom as a soliton of the electrostatic field, if
one likes. Fortunately for the chiral soliton, corrections to the
model go as $1/N_c$ or even as $1/N_c^2$ and can be computed for
many observables. In the Thomas--Fermi model of atoms corrections
to the mean field are of the order of $1/\sqrt{Z}$ and
for that reason are large unless atoms are heavy. 

\begin{figure}[t]
\epsfxsize=14pc 
\epsfbox{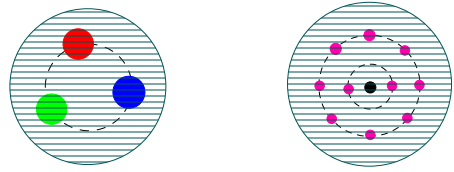} 
\caption{Left: the nucleon is three dynamically-massive quarks bound by the
self-consistent pion field (the `soliton'). The pion field, however, is not elementary
but is made itself of strongly bound quark-antiquark pairs, in accordance with Fig.~10.
Right: an analog -- the Ne atom with electrons bound by the self-consistent electrostatic
field. Fluctuation corrections to the mean field are in both cases about 30\%.}
\vskip -0.4true cm
\end{figure}

There are two instructive limiting cases in the CQSM:  
\begin{itemize}
\item Weak $\pi(x)$ field. In this case the Dirac sea is weakly distorted as
compared to the no-field and thus carries small energy, $E_{\rm sea}\simeq 0$.
Few antiquarks. The valence-quark level is shallow and hence the three 
valence quarks are non-relativistic. In this limit the CQSM becomes very similar
to the constituent quark model remaining, however, relativistic-invariant and well
defined. 
\item Large $\pi(x)$ field. In this case the bound-state level with valence
quarks is so deep that it joins the Dirac sea. The whole nucleon mass
is given by $E_{\rm sea}$ which in its turn can be expanded in the derivatives
of the mean field, the first terms being close to the Skyrme lagrangian.
Therefore, in the limit of large and broad pion field, the model formally reduces
to the Skyrme model.  
\end{itemize}

{\em The truth is in between these two limiting cases.} The self-consistent pion
field in the nucleon turns out to be strong enough to produce a deep
relativistic bound state for valence quarks and a sufficient number of antiquarks,
so that the departure from the non-relativistic constituent quark model
is considerable. At the same time the self-consistent pion field is spatially
not broad enough to justify the use of the Skyrme model which is just a crude
approximation to the reality, although shares with reality some qualitative features. 
The CQSM demystifies the main paradox of the Skyrme model: how can one make a fermion
out of a boson-field soliton. Since the ``soliton'' is nothing but the self-consistent 
pion field that binds quarks, the baryon and fermion number of the whole construction
is equal to the number of quarks one puts on the valence level created by that field:
it is three in the real world with three colors. See Ref.~\cite{DP-Ioffe} for a review.

\section{Nucleons under a microscope with increasing resolution}

\begin{figure}[]
\epsfxsize=20pc 
\epsfbox{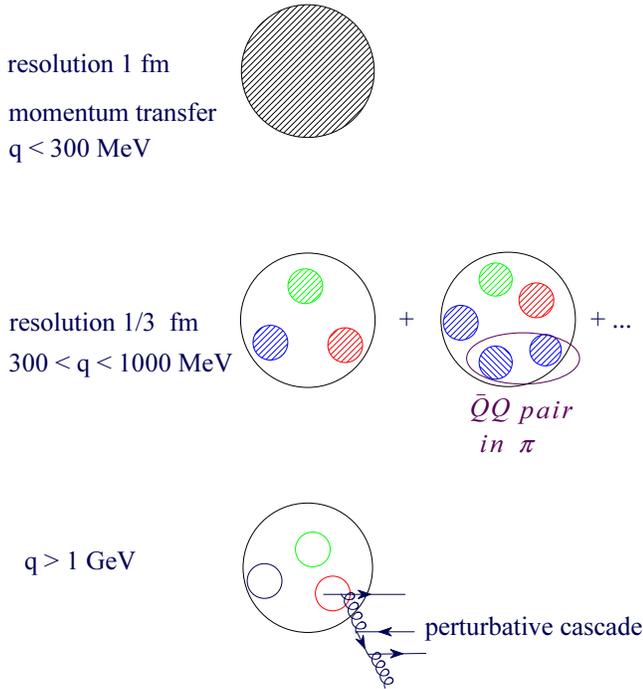} 
\caption{Nucleons under a microscope with increasing resolution.}
\end{figure}

Inelastic scattering of electrons off nucleons is a microscope with which we look into
its interior. The higher the momentum transfer $Q$, the better is the resolution of this 
microscope, see Fig.~12. 

At $Q<300\,{\rm MeV}$ one does not actually discern the internal structure; it is the domain 
of nuclear physics. At $300<Q<1000\,{\rm MeV}$ we see three constituent quarks 
inside the nucleon, but also additional quark-antiquark pairs; mathematically, 
they come out from the distortion of the Dirac sea in Fig.~10. The appropriate quark 
and antiquark distributions have been found in Ref.~\cite{SF}.
In addition, the non-perturbative gluon distribution appears for the first time
at this resolution. First and foremost, it is the glue in the interior of the constituent
quarks that has been responsible for rendering them the mass, i.e. glue from the instanton
fluctuations. It is amusing that these non-perturbative gluons are emitted not
by the vector (chromoelectric) quark current but rather by the quarks' large chromomagnetic
moment, and their distribution has been found by Maxim Polyakov and myself 
to be given by a universal function $(1-x)/x$, see section 7 in Ref.~\cite{inst_at_work}.

At large $Q>1\,{\rm GeV}$ one gets deep inside constituent quarks and starts to see
normal perturbative gluons and more quark-antiquark pairs arising from bremsstrahlung. 
This part of the story is well-known: the perturbative evolution of the parton cascade
gives rise to a small violation of the Bjorken scaling as one goes from moderate 
to very large momentum transfers $Q$, but the basic shape of parton distributions 
serving as the initial condition for perturbative evolution, is determined at moderate 
$Q$ by non-perturbative physics described above. \\

\underline{What is computed in the Chiral Quark Soliton Model} 

\begin{itemize}
\item `Static' characteristics of baryons: masses, magnetic moments, formfactors, 
      coupling constants, $g_A$, etc., see Ref.~\cite{Review}. Important: the large value
      of the nucleon $\sigma$-term explained~\cite{DPPrasz}, as due to the additional 
      $Q\bar Q$ pairs in the nucleon to which the $\sigma$-term is particularly sensitive.   
\item Parton distributions at low virtuality, automatically satisfying the known QCD sum rules~\cite{SF}.
      Explained: a relatively large fraction of antiquarks, large flavor asymmetry of the
      sea $\bar u(x)-\bar d(x)$~\cite{FA}, small fraction of the nucleon spin carried by 
      the quarks' spin~\cite{WY}.   
\item Distribution amplitudes or light-cone quark wave functions in $N$ and $\Delta$~\cite{light-cone}.
      One recovers approximately the old non-relativistic $SU(6)$ wave function but with
      relativistic corrections and the additional $Q\bar Q$ pairs. It is important for
      understanding hard exclusive reactions~\cite{GPV}. 
\item Prediction: large flavor asymmetry in the {\em polarized antiquark} parton distributions
      $\Delta\bar u(x) - \Delta\bar d(x)$~\cite{polar-asym}.  
\item Prediction: peculiar oscillatory behavior of generalized parton distributions~\cite{GPD}.
\end{itemize}

\section{Baryon excitations}

There are excitations related to the fluctuations of the chiral field about its
mean value in the baryons. In the context of the Skyrme model many resonances
were found and identified with the existing ones in Ref.~\cite{HEHW,MK} and
quite recently in Ref.~\cite{Klebanov}. As I said before, the Skyrme model
is too crude, and one expects only qualitative agreement with the Particle Data.
The same work has to be repeated in the CQSM but it has not been done so far. 

There are also low-lying collective excitations related to slow rotation of the
self-consistent chiral field as a whole both in ordinary and flavor
spaces. The result of the quantization of such rotations was first given by
Witten~\cite{Witten}. The following $SU(3)$ multiplets arise as 
rotational states of a chiral soliton: $\left({\bf 8},\frac{1}{2}^+\right), 
\left({\bf 10},\frac{3}{2}^+\right), \left({\bf \overline{10}},\frac{1}{2}^+\right),
\left({\bf 27},\frac{3}{2}^+\right), \left({\bf 27},\frac{1}{2}^+\right)...$
They are ordered by increasing mass, see Fig.~13. The first two (the octet and the decuplet) 
are indeed the lowest baryons states in nature. They are also the only two
that can be composed of three quarks, according to the quantum numbers. However, the
fact that one can manage to obtain the correct quantum numbers of the octet 
and the decuplet combining only three quarks, does not necessarily mean that
they {\em are} made of three quarks only. The difficulties of such an interpretation
have been mentioned in the beginning. In fact we know for thirty years that
the `three-quark' view is simplistic: even baryons from the lowest  
$\left({\bf 8},\frac{1}{2}^+\right),\left({\bf 10},\frac{3}{2}^+\right)$ multiplets
have always an admixture of $Q\bar Q$ pairs, see Fig.~12. 

Therefore, one should not be {\em a priori} confused by the fact that higher-lying 
multiplets cannot be made of three quarks: even the lowest ones are not. A more
important question is where to stop in this list of multiplets. Apparently
for sufficiently high rotational states the rotations become too fast: 
the centrifugal forces will rip the baryon apart. Also the radiation of pions
and kaons by a fast-rotating body is so strong that the widths of
the corresponding resonances blow up~\cite{Regge}. Which precisely rotational excitation
is the last to be observed in nature, is a quantitative question. Actually
one needs to compute their widths in order to make a judgement. If the width
turns out to be in the hundreds of MeV, one can say that this is where the rotational 
sequence ceases to exist. 

\begin{figure}[t]
\epsfxsize=20pc 
\epsfbox{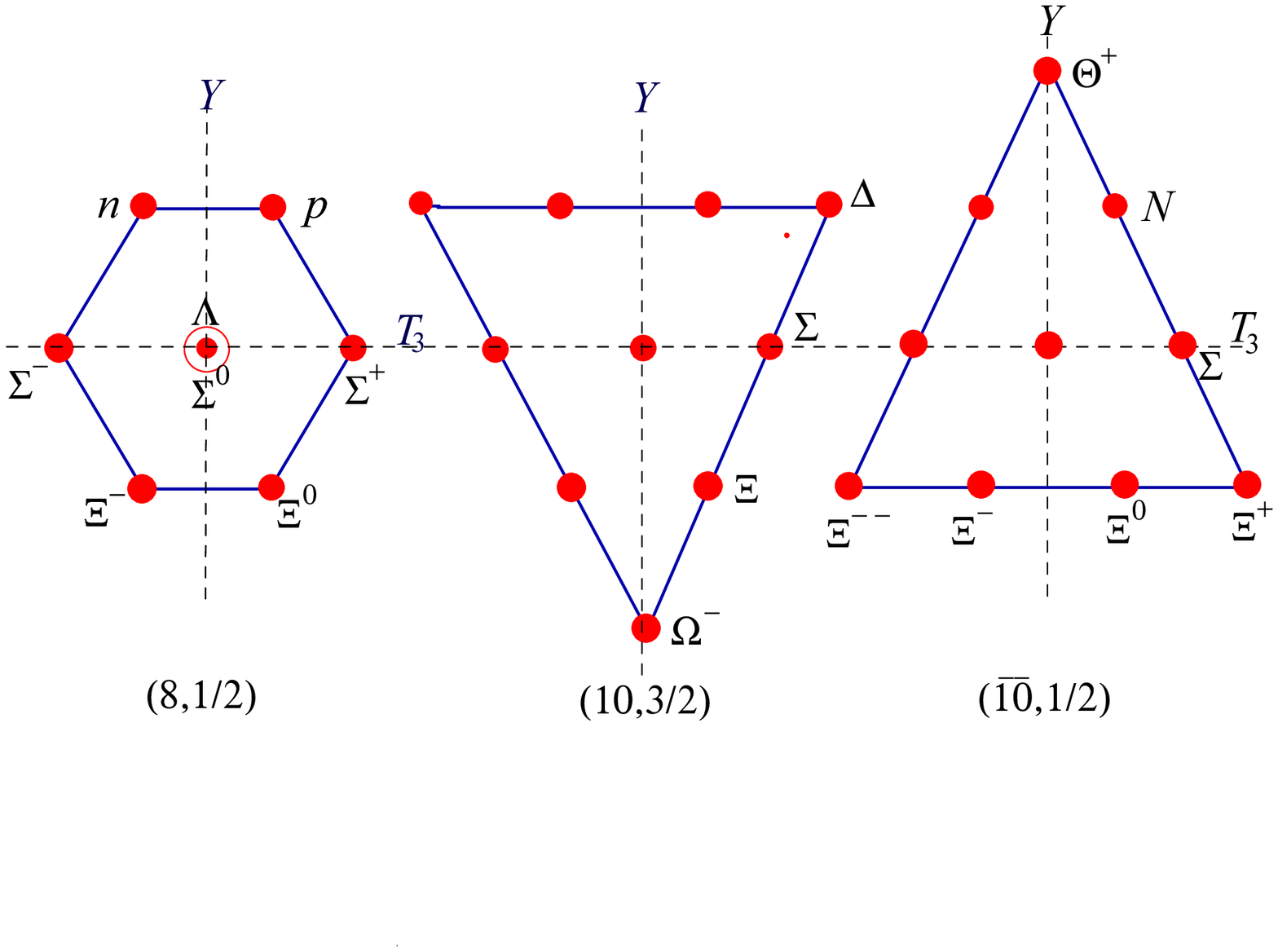} 
\vskip -1.3true cm

\caption{The lowest baryon multiplets which can be interpreted
as rotational states in ordinary and 3-flavor spaces, shown in the $Y-T_3$
axes.}
\vskip -0.3true cm
\end{figure}

An estimate of the width of the lightest member of the antidecuplet, shown at the top 
of the right diagram in Fig.~13, the $\Theta^+$, gave a surprisingly small result:
$\Gamma_{\Theta}<15\,{\rm MeV}$~\cite{97}. This result obtained in the CQSM, immediately 
gave credibility to the existence of the antidecuplet. It should be noted that 
there is no way to obtain a small width in the oversimplified Skyrme model.  

Let us recall our bookkeeping for baryon masses. \\

\noi 
\underline{Baryon masses again} ($M_Q\approx 350\,{\rm MeV}$)
\bea\nn
{\rm mainly\; 3\!-\!quark\; baryons:\;}
{\cal M}&\approx & 3M\; (+\, {\rm strangeness}),\\
\nn
{\rm 5\!-\!quark\; baryons,\;naively:\;}
{\cal M}&\approx & 5M\; (+\, {\rm strangeness})\\
\nn
&\approx & 1900\,{\rm MeV\; for}\; \Theta^+ \; ??\\
\nn
{\rm 5\!-\!quark\; baryons,\; correct:\;}
{\cal M}&\approx & 3M +\frac{1}{\rm baryon\;size}\\
\nn
(+\, {\rm strangeness})&\approx & 1550\,{\rm MeV}\; {\rm for}\; \Theta^+. 
\eea 

In pentaquarks forming the antidecuplet shown on the right of Fig.~13, the additional 
$Q\bar Q$ pair is added in the form of the excitation of the (nearly massless) 
chiral field. Energy penalty would be zero, had not the chiral field been restricted 
to the baryon volume. This is clearly seen from the formulae. The splitting
between the average masses of octet, decuplet and antidecuplet baryons is given
by the two moments of inertia $I_{1,2}$: 
\beq
{\cal M}_{({\bf 10},\,\frac{3}{2})}-{\cal M}_{({\bf 8},\,\frac{1}{2})}
=\frac{3}{2}\frac{1}{I_1},
\la{octdecspl}\eeq
\beq
{\cal M}_{({\bf \at},\,\frac{1}{2})}-{\cal M}_{({\bf 8},\,\frac{1}{2})}
=\frac{3}{2}\frac{1}{I_2},
\la{oct-antidecspl}\eeq
The first, $I_1$, is the moment of inertia with respect to the usual rotations 
of a baryon; it coincides with the moment of inertia with respect
to rotations of the chiral field in the isospin space. 
Numerically, it turns out to be about $(1\,{\rm fm})^{-1}$, a very
reasonable value for a baryon. The second, $I_2$, is the moment of inertia
with respect to rotations mixing `strange' directions of the flavor space.
[In the chiral limit when the current mass $m_s$ is neglected such
rotations are required by the $SU(3)$ symmetry, and are as good as the isospin
rotations.] Numerically, it turns out to be about twice less than $I_1$,
leading to approximately twice larger splitting of the antidecuplet from 
the ground-state octet than the decuplet-octet splitting. 

Important, the splitting \ur{oct-antidecspl} is not twice the constituent mass 
$2M$ but less. One can imagine a large-size baryon: in such a case the moment of 
inertia, roughly $I\sim m<\!r^2\!>$, blows up, meaning that it costs little energy 
to excite the antidecuplet!

\section{Prediction of the $\Theta^+$}

In 1997 Victor Petrov, Maxim Polyakov and I summarized what we knew about the 
antidecuplet. There were two striking features: it had to be a) relatively
light and b) surprisingly narrow for strongly decaying resonances with masses
above $1.5\,{\rm GeV}$. 

On the theoretical side, there were predictions for pentaquarks in the constituent 
quark models~\cite{bag-model} with the lightest one appearing in the range
$1700-1900\,{\rm MeV}$. These models predicted {\em negative}-parity baryons to be 
lighter than positive-parity ones, meaning that they could decay in the $s$-wave
and had to be extremely broad. In contrast to those {\it ad hoc} models, the 
Skyrme model correctly emphasized the importance of the chiral degrees of freedom 
in baryons. It therefore predicted more light pentaquarks, the lightest being 
in the $1400-1700\,{\rm MeV}$ mass range~\cite{Skyrme-predictions}.
However, the Skyrme model was just a qualitative approximation to the reality,
and one faced a difficult task of getting the correct observables for 
ordinary baryons, not to mention the exotic ones. To the best of my knowledge, none 
of the authors attempted to estimate the widths of the antidecuplet baryons, which 
was the crucial thing (had such an attempt been made, it would have yielded a factor 
of 5-15 larger width than in the CQSM). Since there were too many inherent inconsistencies 
inside the models considered, none of the authors seemed to have taken the antidecuplet 
for real. 
 
On the experimental side, from 1960's till early 80's there have been
intensive searches for the exotic $S=+1$ baryons (called $Z$ baryons
in those days) with no convincing results. There have been several
one- and two-star resonances summarized by the Particle Data
Group in the 1986 edition, all in the $1700-1900\,{\rm MeV}$ mass
range. After 1986 the Particle Data Group ceased to mention those resonances as
insufficiently convincing. The 1992 partial wave analysis of the
$KN$ scattering data \cite{HARW1992} concluded that there might be
broad resonances but, if any, they ought to be in a high-mass range, 
see Fig.~14.  

\begin{figure}[t]
\epsfxsize=15pc 
\epsfbox{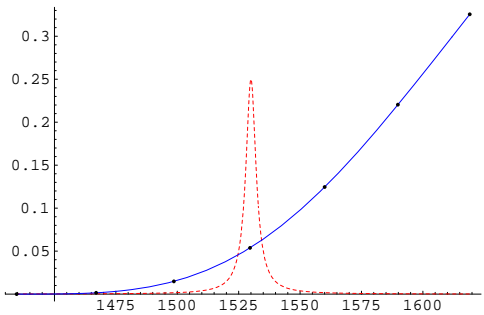} 
\caption{Solid line: modulus squared of the ``exotic'' $KN$ scattering amplitude 
in the $P_{01}$ wave; data points are from the 1992 phase shift analysis summarizing 
the existing data at that time~\cite{HARW1992}. The dashed line is the 1997 
prediction of the $\Theta^+$~\cite{97}, assuming its width is $5\,{\rm MeV}$.
Could a narrow resonance been missed?}
\end{figure}

Despite a heavy pressure from the unsuccessful attempts to find exotic baryons in the past
and despite a general trend to dogmatize the ``three quarks only'' view on baryons, we decided
to publish our findings:\\

\noi {\bf Exotic Anti-Decuplet of Baryons: Prediction from Chiral
Solitons} [Z. Phys. A 359, 305 (1997)]\\ 

\noi {\bf Abstract:} We predict an exotic baryon (having spin $1/2$, isospin $0$ and
strangeness $+1$) with a relatively low mass of about $1530\,{\rm MeV}$ and total width 
of less than $15\,{\rm MeV}$. It seems that this region of masses has avoided thorough
searches in the past. \\

\begin{figure}[t]
\hskip -1.5true cm

\epsfxsize=22pc 
\epsfbox{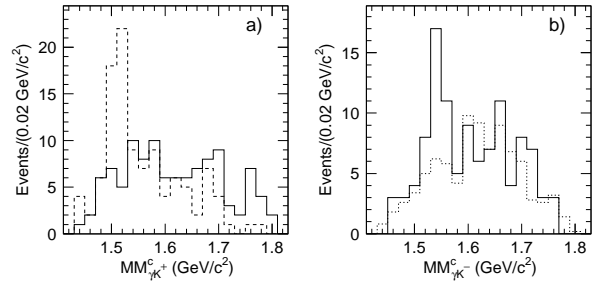}
\vskip -0.6true cm
 
\caption{LEPS experiment~\cite{Osaka}. Left panel, solid histogram: spectrum of $K^-n$ invariant masses                
(no resonances expected); dashed: spectrum of $K^-p$ invariant masses                   
($\Lambda$(1520) expected and seen). Right panel, solid histogram: spectrum of 
$K^+n$ invariant masses ($\Theta^+$ expected); dashed histogram: spectrum of 
$K^+p$ invariant masses. Conclusion by the authors:                                             
$m_{\Theta}=1540\pm 10\,{\rm MeV}$, $\Gamma_\Theta < 25\,{\rm MeV}$, $\sigma = 4.6$.}
\vskip -0.3true cm

\end{figure}

In February 2000 at a conference in Adelaide, Australia, I managed to convince
Takashi Nakano, the spokesman of the LEPS collaboration at SPring-8 near
Osaka, that it is worthwhile to search for this particle in reactions induced
by high-energy gamma quanta. In October 2002 Takashi Nakano reported on the first 
evidence of the new baryon from the $\gamma$C reaction at the PANIC 2002 conference 
in Osaka~\cite{Osaka}. Independently, the DIANA collaboration lead by Anatoly Dolgolenko 
at ITEP, Moscow, looked into the K$^+$Xe bubble chamber data taken at ITEP back in 1986.
They learned about our prediction from a talk given by Polyakov. Rescanning old films 
took more than two years, and in December 2002 the group reported on the observation
of a very narrow $\Theta^+$~\cite{ITEP}. The SPring-8 and ITEP groups did not know
about each other's work, used completely different reactions, and observed the $\Theta^+$
peak using different techniques and final states, see Figs.~15,16. I believe that both
groups can claim credit for the first independent evidence of the $\Theta^+$. 
The next groups knew about these two experiments,
and gave very important confirmation using various reactions and final states, with higher 
statistical significance~\cite{JLab1,JLab2,ELSA,neutrino,HERMES,Protvino,Juelich,Dubna,ZEUS}. 
It should be noted, however, that there have been several experiments where $\Theta^+$ 
has not been seen~\cite{no-Theta-1,no-Theta-2} (I briefly comment on their non-sighting
at the end). Therefore, one is now looking forward to the next tour of dedicated experiments 
with higher statistics, for the issue to be finally resolved.

\begin{figure}[b]
\vskip -0.5true cm

\epsfxsize=16pc 
\epsfbox{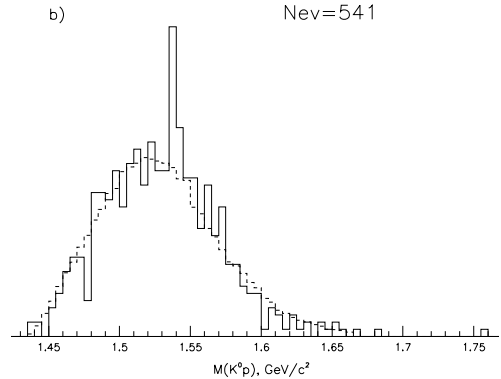} 
\caption{ITEP bubble chamber experiment~\cite{ITEP}: $K^+{\rm Xe}\to \Theta^+\to K^0_Sp$. 
Conclusion by the authors: ${\cal M}_\Theta=1539\pm 2\,{\rm MeV}$,
$\Gamma_\Theta < 9\,{\rm MeV}$, $\sigma=4.4$.}
\end{figure}

The ITEP experiment~\cite{ITEP} being so far the only one where the resonance is formed
directly in the $K^+n$ reaction, provides an additional valuable information. 
First, it gives the most stringent direct restriction on the $\Theta^+$ width, less than
$9\,{\rm MeV}$. Moreover, since the formation cross section is proportional to the
width it can be estimated indirectly from the number of the observed events in the
peak, with the result $\Gamma_\Theta=0.9\pm 0.3\,{\rm MeV}$~\cite{CT} (a conservative 
estimate is less than $3.6\,{\rm MeV}$). Such a narrow width is compatible with other 
indirect estimates based on the fact that $\Theta$ has not been seen in previous 
experiments~\cite{narrow}. In the $K^+n$ reaction, $\Theta^+$ is formed with the lab 
momentum $440\,{\rm MeV}$ and the speed $v_\Theta=0.27\,c$. Taking the $1\,{\rm MeV}$ 
width, one finds that it travels to a distance of $55\,{\rm fm}$ before it decays, 
far beyond the heavy nucleus where it has been produced! Therefore, the second piece 
of information one can extract from the ITEP data is that the $\Theta^+$ must be a compact 
object with weak interaction with the nuclear medium, otherwise it would have been rescattered 
or destroyed on its way through the large Xe nucleus. In particular, the weakly bound 
molecule-type $KN$ bound state can be probably ruled out. 

If it indeed exists, $\Theta^+$ must be not only exotic in quantum numbers but a very 
unusual baryon: extremely narrow, compact and weakly interacting. \\   
 
\underline{Why is it named $\Theta^+$?}\\

After the two groups have announced the evidence of a new exotic baryon, I  
sent out a letter to all parties involved suggesting to name it ``the $\Theta^+$
baryon'', with the following arguments~\cite{letter} :
\begin{itemize}
\item The old name $Z$ used sometimes in the past for exotic candidates is
unfortunate -- first, because it can be confused with the electroweak boson
and second, because the old one- and two-star candidates were broad and in a much 
higher mass range
\item It must be named by an upper-case Greek character, according to the tradition
of naming baryons
\item It must be distinct from anything used before and carry no
associations with bosons
\item The character must exist in LaTeX
\end{itemize}
The only capital Greek character satisfying all criteria is $\Theta$.
It is a nice ``round'' character hinting that it is an isosinglet, like
the $\Omega^-$ hyperon also sitting at the vertex of a big triangle (Fig.~13).
However, in distinction from the $\Omega^-$ which has
been predicted by Gell-Mann to {\em close} the decuplet of {\em known} 
baryons, $\Theta^+$ (if confirmed) {\em opens} the antidecuplet of {\em unknown} 
exotic baryons; some of them might have already been seen but other still await 
observation.

\section{Why is $\Theta^+$ so narrow?}

All experiments giving evidence of the $\Theta$ see that it is narrow,
the most stringent bound being $\Gamma<9\,{\rm MeV}$~\cite{ITEP}.
The indirect estimates~\cite{CT,narrow} show that it can be actually 
as small as $1\,{\rm MeV}$ or even less. If correct, $\Theta$ would 
be the most narrow strongly decaying particle made of light quarks. 
Any theoretical model of $\Theta$ has to explain the unusually small 
width first of all. 

The Chiral Quark Soliton Model from where the story started, gave an
estimate $\Gamma < 15\,{\rm MeV}$~\cite{97} or, to be more precise,
$3.6\,{\rm MeV} < \Gamma < 11.2\,{\rm MeV}$~\cite{DPP04}.
A more careful analysis of the semileptonic hyperon decays (serving as an 
input for the estimate) performed in Ref.~\cite{Rathke} led to $\Gamma < 5\,{\rm MeV}$.
This is not bad but as for now there are big theoretical uncertainties in the estimate.    

Recently, there have been several attempts to explain the narrow width
qualitatively from various 5-quark non-relativistic wave functions of the $\Theta$
or from ideas how such wave functions could look like. I think such attempts are
futile, for several reasons, and a different technique has to be used to answer
this intriguing question.

First, already in the nucleon the constituent quarks are relativistic. Had they 
been non-relativistic, one would face the paradoxes mentioned in the beginning of the paper, 
and in fact many other. One may say: OK, let the three constituent quarks in the nucleon 
be relativistic. But it makes matters even worse: the more relativistic are the quarks,
the less is their ``upper component'' of the Dirac bispinor, meaning there are {\em less} than three quarks 
in the nucleon. Since the number of quarks {\em minus} the number of antiquarks is the conserved 
baryon number, it automatically means that one gets a {\em negative} distribution of 
antiquarks from the relativistic Dirac equation~\cite{SF}. This is nonsense. People who 
have attempted to get parton distributions from the constituent quark models have not 
focused on this paradox but it is an inevitable mathematical consequence of the Dirac equation. 
It is cured by adding the Dirac sea to valence quarks; only then the antiquark distribution
becomes positive-definite and satisfies the general sum rules~\cite{SF}. Thus, a fully 
relativistic description of nucleons is a must if one does not wish to violate
general theorems. 

Second, when one passes from the nucleons to the $\Theta^+$, the situation becomes
even more dramatic. As stressed above, the $\Theta^+$ is relatively light because
the additional $Q\bar Q$ pair is added in the form of the excitation of a
light and strongly bound chiral field. If there have been any doubts that the three
constituent quarks in the nucleon are relativistic, there is no place for such doubts
for pseudoscalars as the quark masses there have to be `eaten up' to zero, since
they are Goldstone particles. Therefore, quarks in the $\Theta^+$ are essentially 
relativistic.

\begin{figure}[t]
\epsfxsize=8pc 
\epsfbox{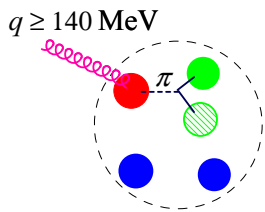}
\caption{Uncertainty principle at work: When one attempts to measure the quark position
in the nucleon to an accuracy better than the pion Compton wave length of $1\,{\rm fm}$ 
one produces a pion, i.e. a new $Q\bar Q$ pair. Hence, the quantum-mechanical description
of baryons with a fixed number of quarks, is senseless.}
\end{figure}

Third, and most important. It has been known since the work of Landau and Peierls (1931)
that the quantum-mechanical wave function description, be it non-relativistic or relativistic, 
fails at the distances of the order of the Compton wave length of the particle.
Measuring the electron position with an accuracy better than $10^{-11}\,{\rm cm}$ 
produces a new electron-positron pair, by the uncertainty principle. One observes
it in the Lamb shift and other radiative corrections. Fortunately, the atom 
size is $10^{-8}\,{\rm cm}$, therefore there is a gap of three orders of magnitude 
where we can successfully apply the Dirac or even the Schr\"odinger equation. 
In baryons, we do not have this luxury. Measuring the quark position with an accuracy 
higher than the {\em pion} Compton wave length of $1\,{\rm fm}$ produces a pion, 
i.e. a new $Q\bar Q$ pair. Therefore, there is no room for the quantum-mechanical wave 
function description of baryons at all. To describe baryons, one needs a quantum field 
theory from the start, with a varying number of $Q\bar Q$ pairs, because of the spontaneous 
chiral symmetry breaking which makes pions light.  

The Chiral Quark Soliton Model provides such a quantum field-theoretic description 
of baryons, and to my knowledge, it is the only one today. This is why it is capable 
to explain the paradoxes of the constituent quark models, and of getting reasonable 
baryon properties, both from the phenomenological and theoretical points of views. 
However, it has a shortcoming: a reference to the large number of colors to justify 
the use of the mean field in baryons. Some day an accurate calculational scheme
for baryons will be developed without referring to the mean field, as it has been
developed for the atoms, but today we have to use the ``Thomas--Fermi approximation"
to understand baryons. 

There is a well known alternative to the
field-theoretic description when one returns to the more intuitive quantum mechanics.
It is the use of the light cone quantization or the infinite momentum frame 
(IMF)~\cite{Brodsky,Wilson}. In that frame, and only in it, there is no production 
and annihilation of particles, so that the light-cone wave function makes sense, 
it does not contradict the Landau--Peierls limitation. There exists a prejudice 
that one uses the IMF only for high energy processes. This is not so: one can use 
it to find the static characteristics of baryons (like the magnetic moments or 
the axial constants). It is just a method where the language of wave functions 
makes sense despite that the number of constituents is not fixed. In the baryon rest frame 
the use of the wave function is contradictory in terms -- it is like using 
coordinates and velocities in describing the hydrogen atom: no respect to the 
uncertainty principle.  

In the IMF, the baryon wave function falls into separate sectors of the Fock space: 
three quarks, five quarks, etc. The general baryon light-cone wave function 
corresponding to the QCSM has been recently derived by Petrov and Polyakov~\cite{light-cone}.
The difference between the ordinary nucleon and the $\Theta^+$ is that the 
nucleon has a three-quark component (but necessarily has also a five-quark component) 
while $\Theta$'s Fock space starts from the five-quark component. In fact, the $\Theta$'s 
wave function is not qualitatively different from the five-quark component of the 
nucleon wave function (Fig.~18). They have similar distributions in longitudinal
and transverse momenta, although their spin-flavor parts are, of course, different~\cite{DPPgA}. \\

\begin{figure}[]
\epsfxsize=20pc 
\epsfbox{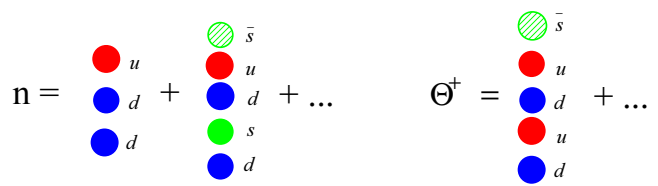}
\caption{In the infinite-momentum frame (and only there!) baryon wave functions are
well defined and fall into separate sectors of the Fock space: 3 quarks, 5 quarks,...
$\Theta$'s wave function is not qualitatively different from the five-quark
component of the nucleon wave function.}
\end{figure}

\noi\underline{$\Theta^+\to nK^+$ decay} \\
\vskip 0.4true cm

Now we finally come to the question how to evaluate the $\Theta$ width. As I stressed
above, in the constituent quark models where quarks are described by some wave function 
in the baryon rest frame, there is no possibility even to ask this question in a   
consistent way, leave alone to answer it. With the light-cone quantization, it is a 
legitimate question that can be answered. 

Let us consider the decay amplitude of the $\Theta$ (or $\Xi_{3/2}$) into an octet baryon 
and a pseudoscalar meson. It is determined by the transition pseudoscalar coupling
$g_{K{\rm N}\Theta}$ (similar to the diagonal $\pi N$ coupling $g_{\pi{\rm NN}}$) which is
related to the transition axial coupling $g_A^{\Theta\to NK}$ (similar to the nucleon 
axial constant $g_A$) by the approximate Goldberger--Treiman relation
\beq
g_{K{\rm N}\Theta}\;\approx \;\frac{g_A^{\Theta\to NK}(m_N+m_\Theta)}{2F_K}
\la{GT}\eeq 
where $F_K\approx F_\pi$ is the kaon decay constant. Hence, it is sufficient 
to evaluate $g_A^{\Theta\to NK}$ as a transition matrix element of the axial charge 
between the $\Theta$ and the nucleon states. In the infinite-momentum frame there can be no 
production and annihilation of quarks, therefore the operator of the axial charge 
does not create or annihilate quarks but only measures the axial charge of 
the existing quarks~\cite{light-cone}. (The same is true for the vector current as well.) 
Thus, the matrix element in question is non-zero only between the pentaquark and the 
{\em five-quark} component of the nucleon!~\cite{DP_mixing} 

To be concrete, let us consider the $\Theta^+\to nK^+$ decay.
In this case, the axial charge has the quantum numbers of the $K^+$ meson, 
$J_{05}^{K^+}=\bar s\gamma_0\gamma_5u$. It annihilates the $u$ quark creating the $s$
quark and annihilates $\bar s$ quarks creating $\bar u$ ones. Correspondingly, there
are basically two processes determining the $\Theta^+\to nK^+$ decay, shown in Fig.~19.
The transition axial constant is given by the normalized matrix element of this
axial charge: 
\beq
g_A^{\Theta^+\to n K^+}=\frac{<\!n^{(5)}|J_{05}^{K^+}|\Theta^+\!>}
{\sqrt{{\cal N}_n^{(3)}+{\cal N}_n^{(5)}+...}\;\;\;\sqrt{{\cal N}_\Theta^{(5)}+...}}
\la{gA}\eeq
where ${\cal N}_n^{(3,5)}$ is the normalization of the 3- and 5-quark components of the neutron
wave function in the infinite-momentum frame $n^{(3)}$ and $n^{(5)}$, respectively; 
${\cal N}_\Theta^{(5)}$ is the same for the $\Theta$; ``..." stand for the omitted 7- and 
higher-quark Fock components. In the numerator, $n^{(5)}$ is normalized to ${\cal N}_n^{(5)}$
whereas in the denominator one has to use the full neutron's normalization. One expects 
that ${\cal N}_n^{(5)}\ll {\cal N}_n^{(3)}$, otherwise the neutron would be a mainly 5-quark baryon.     

\begin{figure}[t]
\epsfxsize=20pc 
\epsfbox{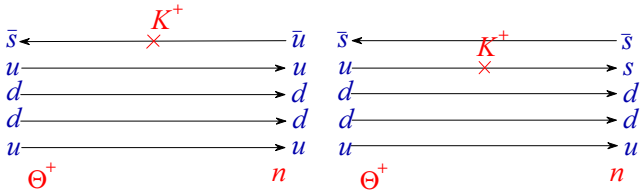}
\caption{Two diagrams determining the $\Theta^+\to nK^+$ decay in the infinite momentum
frame. The cross marked by ``$K^+$'' is where the $K^+$ is emitted: it can either annihilate 
the $\bar s$ quark in the $\Theta$ creating the $\bar u$ quark in the neutron 5-quark component, 
or annihilate the $u$ quark in the $\Theta$ creating the $s$ quark in the neutron. Important,
there is no transition between the $\Theta$ and the 3-quark component of the neutron.}
\end{figure}

Therefore, the transition axial constant $g_A^{\Theta^+\to n K^+}$ is first of all
suppressed by the factor $({\cal N}_n^{(5)}/{\cal N}_n^{(3)})^{1/2}$, that is suppressed 
to the extent the 5-quark component of the neutron is less than its 3-quark component. Additional
suppression comes from the peculiar flavor structure of the neutron's 5-quark component 
where the antiquark is in the flavor-singlet combination with one of the four quarks. 
A detailed calculation is on the way~\cite{DPPgA} but a very crude preliminary
estimate shows that the $\Theta$ width can be as small as $0.7\,{\rm MeV}$.  

It is clear that this suppression of the pseudoscalar transition coupling of the $\Theta$
is quite general, it applies also to scalar, vector, magnetic etc. couplings. It has been
already noticed in the CQSM~\cite{PolRathke}. In fact, all means of exciting the $\Theta^+$
from the nucleon seem to be suppressed, be it via $K$, $K^*$ or $K^*(1430)$ exchanges. 
From this point of view, exciting first some high nucleon resonance with a presumably
large 5-quark component decaying then into the $\Theta$ can be a promising way to
produce it, as suggested by one of the CLAS experiments~\cite{JLab2}. However, the
excitation of a 5-quark nucleon resonance cannot be large either. 

As to the high energy experiments, here the situation with respect to the $\Theta^+$ 
production is probably even worse. At high energies all exchanges with the non-vacuum quantum 
numbers die out and only the gluonic pomeron survives. It may be therefore even more
difficult to excite the 5-quark $\Theta$ by touching a mainly 3-quark nucleon by
soft gluons, than via the meson exchange which is itself small. This could explain 
the present day non-sighting of the $\Theta^+$ in high energy experiments. Probably 
certain kinematical cuts and/or association production should be imposed on the high 
energy data to help disclosing the $\Theta^+$.

\section{Summary}
\vskip 0.3true cm

\noi 1. 93\% of the light baryon masses are due to the Spontaneous Chiral Symmetry
Breaking well explained by instantons. It implies that quarks get a large 
dynamically-generated mass, which inevitably leads to their strong coupling to the chiral 
fields ($\pi, K, \eta$).\\

\noi 2. There is nothing queer in calling baryons solitons -- they are no more solitons 
than atoms, which one may like to call ``the solitons of the self-consistent electrostatic 
field". It is a technical aspect, not a crucial one. The important question is what
forces are responsible for binding quarks.\\

\noi 3. Assuming that the chiral forces are essential in binding quarks together in baryons, 
one gets the lowest multiplets $\left({\bf 8},\frac{1}{2}\right)$, 
$\left({\bf 10},\frac{3}{2}\right)$ and $\left({\overline{\bf 10}},\frac{1}{2}\right)$. 
The predicted $\Theta^+\!\simeq\! uudd\bar s$ is light because it is {\bf not} 
a sum of constituent quark masses but rather a collective excitation of the mean 
chiral field inside baryons.  \\

\noi 4. The non-relativistic wave-function description of an atom
is valid at distances $10^{-8}\,{\rm cm}$, but fails at 
$1/m_ec=10^{-11}\,{\rm cm}$. For baryons, ``$10^{-11}\,{\rm cm}$'' is 
$1\,{\rm fm}$. Relativistic field-theoretic description of baryons is a must.\\

\noi 5. The very narrow width of the $\Theta^+$ can be probably explained. \\

\noi 6. If confirmed, $\Theta^+$ will not only be a new kind of subatomic 
particle but will seriously influence our understanding of how do ordinary nucleons 
``tick'' and what are they ``made of''. \\

Borrowing John Collins' joke: it would be as if a new chemical element between hydrogen 
and helium is discovered. 

\newpage
{\bf Acknowledgments}
\vskip 0.5true cm 

I thank Victor Petrov, Pavel Pobylitsa and Maxim Polyakov for a long-time collaboration
during which the views presented here were formulated, and Yakov \\ Azimov and Mark Strikman
for their comments on the manuscript and useful discussions. I am grateful to the 
Physics Department of the Pennsylvania State University where I have spent a part 
of the sabbatical year, for hospitality and support.

\end{document}